\documentclass[11pt,a4paper]{article}
\usepackage[utf8]{inputenc}
\usepackage{amsmath,amssymb}
\usepackage[english]{babel}
\usepackage{graphicx}
\usepackage{wrapfig}
\usepackage{url}
\usepackage{graphicx}
\usepackage{tabularx}
\usepackage[font={small,it}]{caption}
\usepackage{listings}
\usepackage{scalerel}
\usepackage{calligra}
\DeclareMathAlphabet{\mathcalligra}{T1}{calligra}{m}{n}
\usepackage{float}
\usepackage{amsfonts}
\usepackage{xcolor}
\usepackage{physics}
\usepackage{ulem}
\usepackage{hyperref}
\usepackage{mathtools}
\usepackage{tcolorbox}
\usepackage{soul,xcolor}
\usepackage[toc,page]{appendix}
\usepackage{a4wide}
\usepackage{csquotes}

\usepackage{pdflscape}
\usepackage{parskip}
\usepackage{geometry}
\newenvironment{aleq}
    {\begin{equation}\begin{aligned}}
    {\end{aligned}\end{equation}\ignorespacesafterend}
\usepackage{subfiles}
\graphicspath{{figures/}}
\usepackage{subcaption}
  \usepackage{a4wide}
  \usepackage{latexsym}
  \usepackage{epsf}
  \usepackage{amssymb}
  \usepackage{graphicx}
  \usepackage{amsmath, cite}
  \usepackage{amsmath,amssymb,amsthm}
  \usepackage{verbatim}
  \usepackage{hyperref}
  \usepackage{color}
  \usepackage{mathtools}
  \usepackage{soul,xcolor}

\newcommand{\eq}[1]{(\ref{#1})}

\usepackage{makecell}

\newcommand{\be}{\begin{equation}}
\newcommand{\ee}{\end{equation}}

\begin{document}
\numberwithin{equation}{section}
\begin{flushright}CERN-TH-2025-109\\ IFT-25-062\end{flushright}

\vspace{2.718cm}
\begin{center}

{\LARGE \bf{
Backtracking AdS flux vacua}}\\

\vspace{1cm}

 {\large Fien Apers$^1$, Miguel Montero$^2$, Irene Valenzuela$^{2,3}$}\\
\vspace{1 cm} {\small$^1$ Rudolf Peierls Centre for Theoretical Physics,
Beecroft Building, \\Clarendon Laboratory, Parks Road, University of Oxford, OX1 3PU, UK}\\
{\small$^2$ Instituto de F\'{i}sica Te\'{o}rica IFT-UAM/CSIC,
C/ Nicol\'{a}s Cabrera 13-15, Campus de Cantoblanco, 28049 Madrid, Spain}\\
{\small$^3$CERN, Theoretical Physics Department, 1211 Meyrin, Switzerland}\\
\vspace{0.5 cm} {\small\slshape fien.apers@physics.ox.ac.uk, miguel.montero@csic.es, irene.valenzuela@cern.ch}\\

\vspace{1cm}

{\bf Abstract} \end{center} {We introduce an algorithm (dubbed ``flux backtracking'') to reverse-engineer the brane picture from an AdS flux vacuum. Given an AdS flux vacuum as input, the algorithm outputs a singularity in 10 or 11 dimensions. This singularity has the property that when probed with the appropriate stack of branes (and after taking the near-horizon limit), one recovers the initial AdS vacuum. After testing the procedure on a number of known AdS/CFT pairs, we apply it to AdS flux vacua without known holographic dual, notably the scale-separated DGKT solution. In this case, flux backtracking produces a certain strongly coupled singularity in massive IIA; we conjecture that the worldvolume CFT of D4-branes probing this singularity should be the holographic CFT dual to DGKT (if it exists). Applying the procedure to the DGKT-related scale-separated AdS$_4$ solutions without Romans mass, we find instead a conical and weakly coupled singularity. We also comment on the results and limitations of applying the procedure to KKLT.}

\newpage

\tableofcontents
\section{Introduction}

The AdS/CFT correspondence \cite{Maldacena:1997re, Witten:1998qj, Gubser:1998bc} is an exact match between a conformal field theory (CFT) and an AdS quantum gravity vacuum. The original statement is supported by a vast amount of evidence (e.g.  \cite{Maldacena:1997re, Aharony:1999ti, Klebanov:1998hh, Aharony:2008ug, Jafferis:2012iv, Guarino:2015jca}, which are some of the examples that we will discuss here) in the form of concrete AdS/CFT pairs, where a particular AdS quantum gravity (in practice, an AdS vacuum of string theory) is matched to a known CFT. There is a standard way to produce AdS/CFT pairs from string theory: First, one identifies a suitable stack of branes, perhaps probing a singularity in string theory. If the stack has the right properties, its low-energy worldvolume effective field theory will be a CFT. On the other hand, when this happens, the near-horizon geometry of the brane system provides the dual AdS geometry, thus completing the pair.

Given a brane stack, obtaining the near-horizon AdS is a straightforward (if sometimes cumbersome \cite{Apruzzi:2017nck}) procedure. The \emph{reverse} problem of finding the brane picture given an AdS solution is much harder, and no systematic technique exists, to our knowledge. It is also a much more interesting direction, since there are several known AdS geometries (e.g. \cite{Kachru:2003aw, DeWolfe:2005uu,Cribiori:2021djm, Farakos:2020phe}) with no known holographic dual CFT, including the proposed DGKT vacuum \cite{DeWolfe:2005uu, Camara:2005dc}, which is of independent interest since it exhibits scale separation between the AdS scale and the size of the internal extra dimensions. The field theory dual to DGKT would be extremely interesting, since it must have several exotic properties such as a $N^{9/2}$ scaling of the central charge \cite{Aharony:2008wz, Banks:2006hg, Apers:2022tfm}, \textcolor{black}{a large gap in the spectrum of single-trace operators \cite{Collins:2022nux}}, almost integer conformal dimensions \cite{Conlon:2021cjk,  Apers:2022zjx,Apers:2022tfm, Apers:2022vfp, Quirant:2022fpn, Plauschinn:2022ztd, Andriot:2023fss}, exotic logarithmic terms in the central charge \cite{Bobev:2023dwx} and a very large moduli space lifted by quantum effects \cite{Aharony:2008wz, Montero:2024qtz}. There is no known way to engineer a CFT with these properties; in particular, the rapid scaling of the central charge with $N$ is even larger than $N^3$, which is the largest we know how to engineer from top-down constructions (via M5-branes \cite{Klebanov:1996un, Henningson:1998gx}). Moreover, from the bulk perspective, there is an ongoing debate about whether this proposed four-dimensional supergravity vacuum can be uplifted to a full UV consistent string theory solution (see \cite{Coudarchet:2023mfs} for a recent review).  More recently, similar scale-separated four-dimensional and three-dimensional AdS solutions have been proposed in massless Type IIA using metric fluxes \cite{Cribiori:2021djm} or in massive IIA on G2 orientifolds \cite{Farakos:2020phe}, respectively. Finding the brane picture (if one exists) for either DGKT \cite{DeWolfe:2005uu} or any of these other geometries \cite{Cribiori:2021djm, Farakos:2020phe}  would be a huge leap either towards understanding the dual CFT's or towards making manifest potential UV inconsistencies.

In this short note, we outline a simple procedure, well-known in many concrete examples in holography,  to reverse-engineer a brane picture given an AdS flux compactification. The idea is simple: A typical AdS solution comes out of balancing of curvature and flux terms in an effective action. Fluxes that can be rescaled arbitrarily are precisely those sourced by the probe branes in the dual brane picture; if one sets all of these fluxes to zero, which corresponds to removing all brane charges, the AdS solution disappears; however, one can find instead a running, horizonless solution. Indeed, this is a dynamical cobordism in the sense of \cite{Buratti:2021fiv, Buratti:2021yia, Angius:2022aeq,Blumenhagen:2022mqw,Blumenhagen:2023abk,Mourad:2021roa,Raucci:2022jgw,Basile:2022ypo,Basile:2021vxh,Antonelli:2019nar,Huertas:2023syg,Angius:2023uqk,Angius:2024pqk}, which in examples coincides with the original ten-dimensional geometry probed by the stack of branes. Therefore, to find the brane picture, all one needs to do is find the running solution, and probe it with the stack of branes. Therefore, in some sense, in this paper we are just outlining a systematic procedure to determine particularly interesting examples of dynamical cobordism in flux compactifications. For ease of reference, we dub this procedure ``flux backtracking'', since it allows one to trace back the steps that led to an AdS flux vacuum.

By following the steps above, we are able to recover many known AdS/CFT pairs. For instance, we recover a non-compact Calabi-Yau from compactification on a Sasaki-Einstein manifold, an M-theory orbifold for ABJM or a type I' background from the theories in \cite{Jafferis:2012iv}.  As we will see below, this procedure cannot be applied straightforwardly to all cases, since one needs explicit form for the full, off-shell effective action for arbitrary values of the fluxes, which is not always available. But such an off-shell effective action is available for DGKT \cite{DeWolfe:2005uu} (or the massive IIA vacua of \cite{Guarino:2015jca}), where the brane picture is not known. Applying the procedure to this case results in an unfamiliar singularity in massive IIA. We do not know how to study its low-energy physics, or how to find the worldvolume QFT of D-branes probing it, but this object is a natural candidate for the exotic object providing the brane picture for DGKT.  Similar results are found for the other scale-separated proposals \cite{Cribiori:2021djm, Farakos:2020phe}, although the singularity associated to the vacuum in \cite{Cribiori:2021djm} turns out to be weakly coupled, so it could in principle be analysed using worldsheet techniques.

This note is structured as follows: In Section \ref{s2} we explain the details and limitations of the flux backtracking procedure. In Section \ref{s3} we test the procedure, successfully recovering the known brane picture for several AdS/CFT pairs. Section \ref{sec:noholo} describes the result of flux backtracking in the DGKT solution and other examples where the holographic dual is not known. Finally, Section \ref{s5} contains some concluding remarks.

\section{Flux Backtracking: General Strategy}\label{s2}
Our starting point will be a $d$-dimensional AdS$_d$ vacuum obtained from a flux compactification in string theory. This means that we actually have a higher-dimensional geometry $\text{AdS}_d\times X_n$, where $X_n$ is an internal compactification manifold, such that $d+n=10$ (for perturbative string theory constructions) or $d+n=11$ in the case of M-theory. In many cases of interest, the AdS$_d$ vacuum can be obtained as a solution of a lower $d$-dimensional effective action arising from dimensional reduction on $X_n$,
\begin{equation}S_{d}=\int d^dx\sqrt{-g}\left( \frac{R}{16\pi\, G_d} + \mathcal{L}_{d}(\Phi_i, \vec{n})\right),\label{e34}\end{equation}
where $\Phi_i$ is here a shorthand for the infinite KK tower of fields (of spins 0,1,2) arising from dimensional reduction of the higher-dimensional theory on $X_n$, and $\vec{n}$ is a (properly quantized) vector of integer fluxes threading the compactification manifold $X_n$, which affect the couplings of the lower-dimensional action \cite{Grana:2005jc,Denef:2008wq}. Each choice of $\vec{n}$ is a different low-energy effective field theory, and a different compactification. In examples coming from near-horizon geometries of brane stacks, $\vec{n}$ roughly counts the number of branes (of different kinds) in the stack.

 $\mathcal{L}_{d}(g,\Phi_i, \vec{n})$ is then a very general Lagrangian including kinetic and mass terms for the $\Phi_i$, couplings to the metric $g$, and possibly an infinite tower of higher-derivative couplings arising from dimensional reduction of the higher-dimensional theory. Attempting to find solutions to the equations of motion arising from $\mathcal{L}_{d}(g,\Phi_i)$ is in general a hopeless task, since they involve infinite towers of fields coupled to one another. There are, however, two circumstances in which the problem simplifies and only a finite number of fields becomes relevant:
\begin{itemize}
\item When there is a \emph{consistent truncation}, it is possible to truncate to zero all but finitely many of the $\Phi_i$ and still solve the equations of motion. This usually happens due to supersymmetry of the low-energy effective action \cite{Connes2004}, and it is the case in e.g. the AdS$_5\times S^5$ compactification of IIB string theory \cite{Maldacena:1997re}, where only the lowest modes of the KK tower are switched on, in spite of the fact that they couple strongly to every other KK mode (since the solution is not scale-separated).
\item When the AdS$_d$ solution is \emph{scale-separated}, there is a large gap between the (finitely many) lightest $\Phi_i$ and the rest (see \cite{Coudarchet:2023mfs} for a review). In this case, one can integrate out all heavy fields to obtain the low-energy effective theory including only a finite number of light fields. However, this is typically impossible or extremely hard in realistic string theory compactifications. Instead, if the scale separation is parametrically large,  the heavy fields are expected to have a very small effect on the light ones at low energies, so that one can in principle ignore them and consider the $d$-dimensional effective field theory truncated to the light fields.\footnote{When we obtain a vacuum from simply truncating the theory to the light fields and ignore the effect of the heavy fields, we say that the vacuum does not have a full top-down description in string theory, since it is yet to be proven that such effective theory is indeed the low-energy limit of the full theory after properly integrating out all the heavy fields (including exotic stuff like degrees of freedom at singularities, higher order supersymmetry breaking effects, etc). At the moment, all proposed scale-separated AdS vacua in string theory have been obtained via this truncation, and this is why it is under debate whether they can be lifted to vacua of the full theory.} 
In particular, the DGKT solution of \cite{DeWolfe:2005uu} is famously scale-separated and the low-energy effective field theory of the light fields is well-known.
\end{itemize}
In both these cases, one can further truncate to zero all spin-1, and spin-2 fields, and look for solutions involving the scalars $\phi^i$ only. At the two-derivative level, the action for the scalars simply looks like
\begin{equation} \mathcal{L}_{\text{eff}}^{\text{Scalars}}= \frac12 G_{ij}(\phi^i,\vec{n})  d\phi^i\wedge*d\phi^j- V(\phi^i,\vec{n}),\end{equation}
where $V(\phi^i)$ is a scalar potential. If there is a critical point of $V(\phi^i,\vec{n})$ with $V(\phi^i)<0$, then \eq{e34} admits an AdS$_d$ solution, with cosmological constant given by \begin{equation}\Lambda= 8\pi\, G_{d}\, V_{\text{min}}(\vec{n}).\end{equation} Since $V_{\text{min}}(\vec{n})$ depends on $\vec{n}$, we get a one-parameter family of AdS vacua; the usual dependence is such that $\Lambda\rightarrow0$ as the fluxes get very large.

\begin{figure}[htb!]
\begin{center}
\includegraphics[scale = 0.32]{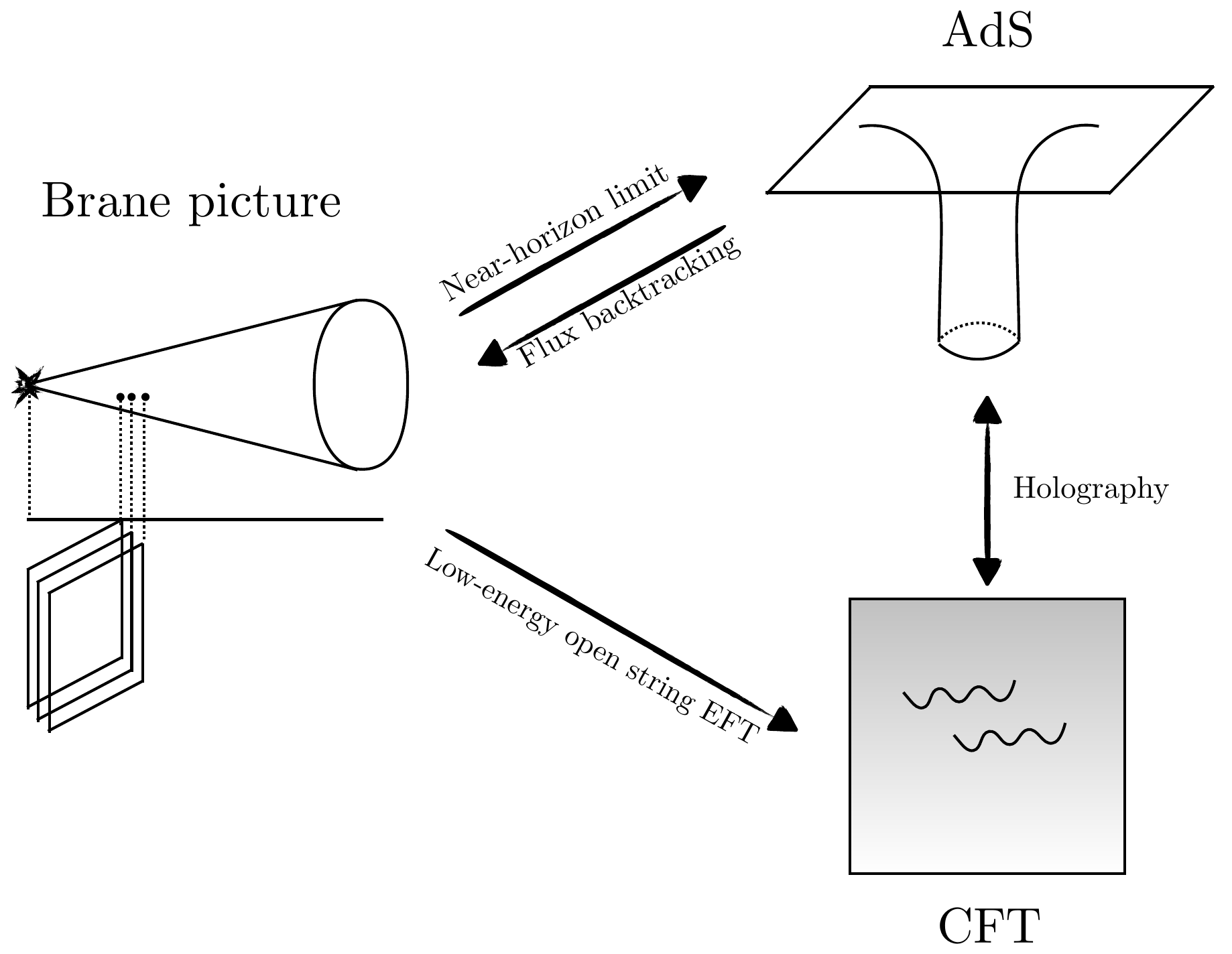}
\caption{Flux backtracking is the inverse operation of taking a near-horizon limit of a brane stack probing a singularity (which produces an AdS geometry under favorable circumstances). It may be of interest in cases where the brane picture and/or the holographic dual are unknown, such as in DGKT or the vacua in \cite{Cribiori:2021djm}.}
\label{f0}
\end{center}
\end{figure}

Our goal is to find out what is the precise geometry that has to be probed by branes, which source the fluxes $\vec{n}$, in order to produce the given AdS$_d$ family of flux vacua as their near-horizon geometry.  In other words, to backtrack the flux compactification in this way, we need to get rid of the flux; so we simply set $\vec{n}=0$ in the above. In this case, the scalar potential $V(\phi^i,\vec{n})$ does \emph{not} have a critical point and hence there is no AdS solution. Instead, in this case one can obtain a running solution, where the metric takes the form 
\begin{equation}\label{eq:DW}  ds_d^2 = dr^2 + e^{2A(r)} ds_{d-1,\text{flat}}^2\end{equation}
where $ds_{d-1,\text{flat}}$ is a flat metric, and where the profiles of all scalar fields $\Phi_i$ depend only on $r$. When $A(r)$ is a linear function of $r$, this metric \eqref{eq:DW} describes (the Poincar\'{e} patch) of AdS$_d$ \cite{Nastase:2007kj}. But when no AdS solution exist, one has more general profiles, similar to the dynamical cobordisms of \cite{Buratti:2021fiv, Buratti:2021yia, Angius:2022aeq}. The resulting running solution with $\vec{n}=0$ is the geometry we are looking for.  A general lesson from those works is that one should expect $A$ to become singular at a particular value of $r$, where the geometry ends in a singularity. Putting back the branes that source the fluxes $\vec{n}$ on the geometry on top of this singularity recovers the original AdS solution we started with, as illustrated in Figure \ref{f0}.

In the next Section we will illustrate the flux backtracking procedure in some very well-known examples of the AdS/CFT correspondence, to check that it produces meaningful answers. We close this Section with some general issues that will appear when applying the procedure: \begin{itemize}
\item The process of flux backtracking is not unique except in very simple cases. In general, it can be done in several ways, corresponding to the different ways there may be to obtain a brane picture for a given AdS vacuum. For instance, the $\text{AdS}_3\times S^3\times T^4$ vacuum dual to the D1-D5 SCFT \cite{Strominger:1996sh} is supported by two fluxes $\vec{n}=(n_1,n_2)$, sourced by D1- and D5-branes respectively. One can get the vacuum by probing the singular geometry sourced by $n_1$ D1-branes with D5-branes, or by probing with D1-branes the \emph{different} singularity sourced by $n_2$ D5-branes. Consequently, there are at least two different ways to implement the flux backtracking procedure.
\item The output of flux backtracking is a singular configuration which is, topologically, a boundary for the starting compactification on $X_n$ -- it is an example of the defects predicted in \cite{McNamara:2019rup, Buratti:2021yia, Angius:2022aeq}. In principle, there are infinitely many ways to do this; for instance, one could consider a cobordism to a different compactification manifold $Y_n$, or change the way that tadpoles are satisfied in $X_n$ by adding additional branes to the compactification, etc. All of these operations change the effective Lagrangian $ \mathcal{L}_{\text{eff}}(g,\Phi_i, \vec{n})$. In this paper, we restrict to flux backtracking within a given effective theory, meaning that the fields in $\mathcal{L}_{\text{eff}}^{\text{Scalars}}$ are all the same for any value of $\vec{n}$. As a result, we do not allow for more exotic cobordisms, and the only fluxes in $\vec{n}$ that we can change freely are those that do not appear in tadpoles of the compactification. It would be interesting to lift these restrictions in the future.

\item When the scalar potential $V(\phi^i,\vec{n})$ is obtained via a consistent truncation, one must be sure that this truncation remains consistent for all $\vec{n}$, including $\vec{n}=0$; otherwise, the $\vec{n}=0$ running solution might require one to leave the truncation, and an extended problem should be considered.
\end{itemize}

Related to the second point above, when testing the flux backreaction procedure in concrete examples, we will find several instances where the lower $d$-dimensional scalar potential is not known -- instead, only the AdS solution is known, and it was obtained directly from the higher-dimensional supergravity. Absent the scalar potential, we are forced to do some informed guesswork about which scalars are relevant and run once the fluxes are switched off. We get the right result because in these examples we know the brane picture anyway; but for any interesting example where the brane picture is not known, we must be sure to include all relevant scalars. This will be a potential issue in some of the examples in \cite{Apruzzi:2017nck}, but importantly, it is \emph{not} an issue for the DGKT solution, precisely because it is supposed to be scale-separated \cite{DeWolfe:2005uu}. As described above, in this case there is an honest lower-dimensional EFT with a finite number of scalar fields, and as long as these are all taken into account, the description is complete.

\section{Vacua with holographic dual}\label{s3}

We are now ready to test the flux backtracking procedure against known holographic examples. We will start with the simplest case where the relevant effective field theory contains a single scalar. As in Section \ref{s2}, we consider a $d$-dimensional action with a scalar potential \( V(\phi) \), expressed as
\begin{equation}
    S_d = \int d^dx \sqrt{-G_d} \left(\frac{1}{2} \mathcal{R}_{d} - \frac{1}{2} G_{ij}(\partial \phi^i)(\partial \phi^j) - V(\phi) \right),
\end{equation}
where $G_{ij}$ is a metric on field space. After removing the flux $\vec{n}$, we will be looking for solutions of the form \begin{equation}\label{flowmetric}
    ds_d^2 = dr^2 + e^{2A(r)} ds_{d-1}^2, \quad \phi_i = \phi_i(r),
\end{equation}
for the metric and scalar field, where just like in Section \ref{s2} $ ds_{d-1}^2$ is a flat metric. This ansatz is precisely what is considered in the study of supersymmetric domain wall solutions \cite{Skenderis:1999mm}, and we will benefit from this fact. Substituting this ansatz into the action yields
\begin{equation}
    S_d = \int d^dx \, e^{(d-1)A} \left(\frac{1}{2} (d-1)(d-2) A'^2 - \frac{1}{2}G_{ij}(\phi^i)' (\phi^j)' - V(\phi) \right).
\end{equation}
From this, we derive the second-order equations of motion\footnote{Assuming the field space metric has vanishing Christoffel symbols, which is always true in the one-dimensional case under consideration here.}:
\begin{align}\label{eoms}
    (\phi^i)'' + (d-1)(\phi^i)' A' &= \partial_{\phi^i} V(\phi), \\
    (d-2)\left[2A'' + (d-1)A'^2\right] &=  - G_{ij}(\phi^i)' (\phi^j)'-2V(\phi),\\
    (d-2)(d-1) A'^2&=G_{ij}(\phi^i)' (\phi^j)'-2V(\phi).
\end{align}

To simplify the analysis, following the techniques from supersymmetric domain walls \cite{Skenderis:1999mm} we aim to recast these into first-order flow equations. A good choice is
\begin{align}\label{eq:flowequations}
    \frac{d\phi^i}{dr} &= \alpha G^{ij}\frac{dP}{d\phi^j}, \quad \frac{dA}{dr} = \beta P,
\end{align}
where \( \alpha \) and \( \beta \) are constants 
\textcolor{black}{given by
\begin{equation}
    \alpha = \pm 2 (d-2), \quad \beta = \mp 2
\label{flafla}\end{equation}
and \( P \) is a potential function. The change of coordinates $r\rightarrow-r$ flips the sign choice in \eqref{flafla}, so without loss of generality we can take the positive sign for $\alpha$. For these first-order equations to be consistent with the second-order equations of motion \eqref{eoms}, we  specify that \( P \) relates to the scalar potential \( V(\phi) \) by
\begin{equation}\label{scalarpotential}
    V = \frac{1}{2} \alpha^2 \left[\sum_i \left( \frac{dP}{d\phi_i} \right)^2 - \frac{(d-1)}{(d-2)} P^2\right].
\end{equation}}
Under these conditions, solutions to the first-order flow equations automatically satisfy the second-order equations of motion in \eqref{eoms} \cite{Skenderis:1999mm}. \textcolor{black}{Such a potential function $P$ is generally present in supersymmetric frameworks, but it can also arise in contexts involving fake supersymmetry \cite{Freedman:2003ax, DiazDorronsoro:2016rrz}.} We are now ready to implement flux backtracking. 
We will be assuming that the $d$-dimensional effective field theory arises from a higher $d+n$ dimensional theory, such that the metric can be uplifted to something of the form 
 \begin{aleq}
     ds_{d+n}^2 = f(\phi_i) \left(dr^2 + e^{2A(r)} ds_{d-1}^2\right) + g(\phi_i) ds_n^2,
 \end{aleq}
 for some moduli-dependent functions $f$ and $g$. Here, $ds^2_n$ is the metric of the internal compactification $X_n$, which remains the same whether we turn fluxes on or off in all examples considered in this paper (but this will not be the case in more general examples). 

 In general, it will not be easy to determine the number of supercharges preserved by the running solution obtained from flux backtracking a given vacuum. In this paper we have focused on running solutions to the BPS flow equations \eqref{eq:flowequations}, so they will preserve at least half of the supercharges of the original vacuum. However, they may preserve more\footnote{For 4d $\mathcal{N}=1$ vacua, the solution preserves so little supersymmetry that quantum corrections may even break it completely (see e.g. \cite{Montero:2024qtz}).} (as e.g. in the flux backtracking of the $AdS_5\times S^5$ IIB vacuum discussed in Section \ref{subsection:FR}).
 
Finally, we wish to provide a word of caution about low codimension cases. For the picture of a stack of branes probing a singularity to give a field theory description of the system, it is important that bulk low-energy modes decouple from the degrees of freedom described by the brane stack+singularity system. This happens automatically when the codimension of the singularity and/or brane probes is higher than two. Otherwise, the question of bulk mode decoupling is subtler and needs to be analysed on a case by case basis. If bulk modes do not decouple, spin two or higher states may remain in the low-energy theory, thereby spoiling a pure field theoretical description (which can only involve fundamental fields of spin $\leq$ 1). Hence, even if the singularity+branes do describe the system in some sense, we will not be able to use them to learn anything about the system that we did not understand from the original gravity solution. Fortunately, all examples considered in this paper will be of higher codimension when viewed from the 10 or 11 dimensional perspective.
 
 \subsection{Freund-Rubin vacua}\label{subsection:FR}
 We will first consider Freund-Rubin vacua, which are arguably the simplest class of holographic AdS vacua in string theory \cite{Freund:1980xh,Duff:1986hr,Maldacena:1997re}. 
 In this case, the \( (d+n) \)-dimensional action includes contributions from both the curvature of the compact space and a flux threading the \( n \) internal dimensions. The action is given by
\begin{equation}
    S_{d+n} = \int d^{d+n}x \sqrt{-G_{d+n}} \left( \frac{1}{2}\mathcal{R}_{d+n} - \frac{1}{2n!} F_{M_1 \dots M_n} F^{M_1 \dots M_n} \right),
\end{equation}
following the notation in \cite{VanRiet:2023pnx}. In a holographic context, the flux threading the internal dimensions is replaced by D-brane domain walls, with an AdS vacuum emerging as their near-horizon limit. To determine the geometry experienced by these branes, we remove the flux term, leaving a pure Einstein-Hilbert action,
\begin{equation}
    S_{d+n} = \int d^{d+n}x \sqrt{-G_{d+n}}\frac{1}{2} \mathcal{R}_{d+n}.
\end{equation}
Upon dimensional reduction, we obtain
\begin{equation}
    S_d = \int d^dx \sqrt{-G_d} \left(\frac{1}{2} \mathcal{R}_{d} - \frac{1}{2} (\partial \phi)^2 - V(\phi) \right),
\end{equation}
where \( \phi \) is the canonically normalized volume modulus,
\begin{equation}
    \phi = \sqrt{\frac{d+n-2}{n(d-2)}} \ln \mathcal{V},
\end{equation}
defined in terms of the internal volume \( \mathcal{V} \) in string units. The potential for \( \phi \), sourced by the curvature term, reads
\begin{equation}\label{curvaturepotential}
    V(\phi) = -C e^{-2\frac{\sqrt{n+d-2}}{\sqrt{n(d-2)}}\phi},
\end{equation}
where $C$ is a positive constant.
This curvature potential can be rewritten in the form of \eqref{scalarpotential} by defining
\begin{equation}
    P = p e^{-\frac{\sqrt{n+d-2}}{\sqrt{n(d-2)}}\phi}\, .
\end{equation}
where $p^2 = \frac{nC}{2(n-1)(d-2)}$.
Using the ansatz for the metric in \eqref{flowmetric}, the flow equations are solved by
\begin{align}
    \phi(r) &= \sqrt{\frac{n(d-2)}{n+d-2}} \ln r, \quad A(r) = \frac{n}{n+d-2} \ln r,
\end{align}
up to constants. Uplifting the resulting metric to \( d+n \) dimensions yields
\begin{align}
    ds_{d+n}^2 &= \mathcal{V}^{-\frac{2}{d-2}} ds_d^2 + \mathcal{V}^{\frac{2}{n}} ds_n^2 \\
    &= r^{-\frac{2n}{n+d-2}} \left(dr^2 + r^{\frac{2n}{n+d-2}}ds_{d-1}^2\right) + r^{\frac{2(d-2)}{d+n-2}}ds_n^2 \\
    &= dy^2 + ds_{d-1}^2 + y^2 ds_n^2,
\end{align}
where \( y \sim r^{\frac{d-2}{d+n-2}} \) defines a new radial coordinate. This metric describes a conical singularity, implying that all these Freund-Rubin vacua can be described as the near-horizon geometry of the D-branes, that source the flux, probing a conical singularity. This is a well-known fact of Freund-Rubin AdS vacua \cite{Acharya:1998db,Morrison:1998cs}.  Therefore, the flux-backtracking procedure successfully recovers the geometry that must be probed by the branes in this case.

\subsubsection{Example: AdS\texorpdfstring{$_5 \times S^5$}{5xS5} and Sasaki-Einstein compactifications}

The simplest example are the AdS$_5 \times S^5$ vacua in IIB string theory, where a five-form flux \( F_5 \) threads an internal sphere \( S^5 \) \cite{Maldacena:1997re}. To find the brane picture, the $N$ units of five-form flux should be replaced into $N$ D3-branes. Following the results that we obtained above using the flux backtracking method, these D3-branes should probe the following geometry,
\begin{equation}
    ds_{10}^2 = dy^2 + ds_4^2 + y^2 ds_{S^5}^2,
\end{equation}
which is just the 10-dimensional flat space metric. Indeed, it is well known that the near-horizon geometry of a stack of D3-branes in flat space corresponds to AdS$_5 \times S^5$, so the flux backtracking procedure has yielded the correct answer. Moreover, the central charge of the CFT scales as $N^2$, which is precisely the result that one would get from counting perturbative degrees of freedom for a stack of $N$ D3-branes. Since the string coupling does not run in this solution, it can be taken as very small in the solution, so that the worldsheet techniques to count degrees of freedom remain valid (we will see that this is not always the case in more complicated examples).

This solution can be generalized e.g. for the Klebanov-Witten vacua \cite{Klebanov:1998hh} of the form AdS$_5 \times X^5$, where \( X^5 \) is an Einstein space. Using the backtracking flux procedure, the metric then becomes
\begin{equation}
    ds_{10}^2 = dy^2 + ds_4^2 + y^2 ds_{X^5}^2,
\end{equation}
with \( X^5 \) being the base of an actual conical singularity. This geometry is no longer flat space, but indeed corresponds to a conical Calabi-Yau whose base is the Sasaki-Einstein space $X^5$, as expected.  A stack of $N$ D3-branes probing such conical singularity yields the AdS$_5\times X^5$ vacuum as their near-horizon geometry. We can also apply the same logic to any Sasaki-Einstein compactification, yielding the appropriate conical singularity.

\subsubsection{Example: AdS\texorpdfstring{$_4 \times S^7$ }{4xs7}and AdS\texorpdfstring{$_7 \times S^4$ }{7xs4}}

Other well-known examples of Freund-Rubin vacua are the AdS$_4 \times S^7$ and AdS$_7 \times S^4$ solutions of M-theory \cite{Maldacena:1997re}. In these cases, the higher-dimensional spheres are supported by \( N \) units of the seven-form flux \( F_7 \) and four-form flux \( F_4 \) respectively. We can then replace these fluxes by the corresponding stack of \( N \) M2-brane and M5-brane domain walls, respectively, and solve the above flow equations. The result is that these branes should probe an eleven-dimensional flat spacetime with metrics
\begin{equation}
    ds_{11}^2 = dy^2 + ds_3^2 + y^2 ds_{S^7}^2, \quad \text{and} \quad ds_{11}^2 = dy^2 + ds_6^2 + y^2 ds_{S^4}^2.
\end{equation}
recovering the brane picture of these AdS vacua. Here, the central charge grows as $N^{3/2}$ and $N^3$ respectively, corresponding to a stack of $N$ M2-branes and a stack of $N$ M5-branes, respectively.

\subsection{ABJM theories in IIA string theory}\label{ABJM}
Let us move on now to a slightly more complicated family of vacua: the ABJM vacua \cite{Aharony:2008ug}. The ABJM field theories arise as the worldvolume description of $N$ M2-branes placed at a $\mathbb{C}^4/\mathbb{Z}_k$ orbifold singularity, rather than flat space. In the strong coupling limit, where $N\gg k^5$, these are dual to M-theory on AdS$_4 \times S^7/\mathbb{Z}_k$. In the 't Hooft limit, though, where $N, k\rightarrow \infty$ with $\lambda = N/k$ fixed but small, they are better described by Type IIA  on AdS$_4 \times \mathbb{C}P^3$. In this section, we check whether our method recovers the aforementioned orbifold singularity starting from the bulk supergravity description.

We start from type IIA string theory with $N$ units of $F_6$-flux, $k$ units of $F_2$-flux and internal curvature. We remove the contributions from the $F_6$-flux, dual to the M2-branes, from the effective potential and consider the resulting flow generated by $F_2$-flux and internal curvature. In Appendix \ref{app:IIA} we perform the general dimensional reduction of Type IIA and the derivation of the superpotential $P$ for the flux-induced scalar potential. Particularizing it to the case of ABJM, we obtain that the four-dimensional scalar potential is given by
\textcolor{black}{\begin{aleq}
    V = A_R \frac{1}{us^2} + A_{F_2} \frac{u}{s^4} + \text{local\ contributions}
\end{aleq}}
where the first contribution is from the curvature of the internal space and the second one corresponds to the $F_2$-flux. The functions $A_R$ and $A_{F_2}$ are independent of the non-compact moduli so they will not play any role in our discussion.
The universal moduli $u,s$ are related to the overall volume $\mathcal{V}$ and the 10-dimensional string dilaton $\phi$ as follows,
\begin{aleq}
\label{us}
   \mathcal{V} = u^{3} \ \quad e^{-2\phi} = s^{2} u^{-3}.
\end{aleq}
Recall that this is not a proper effective field theory since there is no scale separation; however, thanks to supersymmetry, we can consider a truncation of the theory to the sub-sector of zero modes, which will suffice for our purposes.

Using \eqref{PIIA}, the superpotential $P$ is given by
\begin{aleq}
    P = c_R u^{-\frac{1}{2}} s^{-1} +  c_{F_2} u^{\frac{1}{2}}s^{-2},
\end{aleq}
where we again do not specify the value of the constants as we only need the moduli dependence. 

We find that the flow equations \eqref{eq:flowequations} are solved by
\begin{aleq}
    s(r) \sim r^{2/3}, \quad u(r) \sim r^{2/3}, \quad A(r) = \frac{7}{9} \ln r.
\end{aleq}
The ten-dimensional uplift of the domain wall metric is then given by
\begin{aleq}\label{abjm1}
    ds_{10}^2 &= [s(r)]^{-2} \left(dr^2 + e^{2A(r)}ds_3^2\right) + [u(r)]ds_{\mathbb{C}P^3}^2
    &= dy^2 + y^{2/3}ds_3^2 + y^2ds_{\mathbb{C}P^3}^2,
\end{aleq}
where we have defined $y=r^{1/3}$. This metric represents a deformed conical singularity with a running string coupling
\begin{aleq}\label{abjm3}
    g_s(y) = e^{\phi(y)} =  u^{3/2}(y)s^{-1}(y) =y.
\end{aleq}
When uplifting to eleven dimensions, with $z$ parametrizing the M-theory circle, we have
\begin{aleq}\label{abjm2}
    ds_{11}^2 &= e^{-2\phi(y)/3}ds_{10}^2 + e^{4\phi(y)/3} dz^2\\
    &= d\Tilde{y}^2 + ds_3^3 + \Tilde{y}^2 ds_{S^7/\mathbb{Z}_k},
\end{aleq}
redefining $\Tilde{y} = y^{2/3}$. Therefore, we indeed recover the conical orbifold singularity probed by the stack of M2-branes in the brane picture.
 In conclusion, the solution goes to strong coupling for large $\tilde{y}$, matching the M-theory on $S^7/\mathbb{Z}_k$ description. In the opposite limit $\tilde{y}\rightarrow 0$, the string coupling decreases as we approach the brane stack, so that the system can be analyzed via a D2-brane picture \cite{Aganagic:2009zk}. The central charge of the IR CFT grows as $N^{3/2}$ times a $k$-dependent function \cite{Aharony:2008ug}, generalizing the result of the previous example of $AdS_4\times S^7$. As it is well-known, the tree level $N^2$ naive expectation from a stack of $N$ D2-branes is reduced due to strong coupling effects, but still upper bounded by the perturbative result of the UV worldvolume theory of the D2's since the singularity is weakly coupled.

\subsection{AdS vacua in massive IIA string theory}

Let us next investigate AdS vacua arising from massive Type IIA. We are interested in these vacua as they also include Romans mass, which plays a key role in the DGKT vacuum that we aim to analyse in Section \ref{sec:noholo}. We will see that the flux backtracking procedure provides the correct result in these vacua with Romans mass.

We first consider the singularity probed by D2-branes in the Guarino-Jafferis-Varela set-up \cite{Guarino:2015jca}. This is a squashed AdS$_4 \times S^6$ background in massive type IIA, with dual super-Chern-Simons-matter theories that have a $SU(N)$ gauge group and level $m$ equal to the Romans mass. The CFT lives on the D2-branes which have Romans-mass induced Chern-Simons coupling, which deforms the D2-brane near-horizon geometry.

The lower-dimensional potential of the bulk theory will receive contributions from the $F_6$-flux, the Romans mass and the curvature (see \cite{Guarino:2015qaa}). In order to find the geometry probed by the D2-branes that source the $F_6$-flux, we need to find the running solution to the potential upon setting to zero the $F_6$-flux. Using the result for the effective potential in \eqref{VIIA} of Appendix \ref{app:IIA}, we obtain that the potential driving this running solution is given by
\begin{equation}
    V= A_R \frac{1}{us^2} + A_{F_0} \frac{u^3}{s^4} + \text{local \ contributions} \ \rightarrow P= c_R u^{-\frac{1}{2}}s^{-1} + c_{F_0} u^{\frac{3}{2}}s^{-2},
\end{equation}
where again the moduli $u,s$ are related to the overall volume and the dilaton as in \eqref{us}. The flow generated by the Romans mass and the curvature is solved by
\begin{aleq}
    s(r) \sim r^{4/5}, \quad u(r) \sim r^{2/5}, \quad A(r) = \frac{19}{25} \ln r,
\end{aleq}
and this corresponds to the following ten-dimensional metric,
\begin{aleq}\label{eq:mass_result}
    ds_{10}^2 &= [s(r)]^{-2} \left(dr^2 + e^{2A(r)}ds_3^2\right) + [u(r)]ds_{\Tilde{S}^6}^2
    &= dy^2 + y^{-2/5}ds_3^2 + y^2ds_{\Tilde{S}^6}^2,
\end{aleq}
with $y = r^{1/5}$ and running coupling $  g_s(y) \sim y^{-1}$. This describes a deformed conical singularity, as perhaps could have been expected from the fact that massive IIA does not admit 10-dimensional flat space as a solution. The central charge is known to grow as $N^\frac53$. Since the singularity is strongly coupled, unlike in previous examples, we do not expect to be able to use worldsheet techniques to derive this result. 

The result \eqref{eq:mass_result} is, to our knowledge, the first attempt in the literature at constructing the massive IIA configuration dual to the theories in \cite{Guarino:2015jca}. However, we emphasize that the result should be taken with a grain of salt. As explained in Section \ref{s2}, the simple flux backtracing procedure that we perform relies crucially on us including all light scalars that are relevant to the problem. Since we only started from a four-dimensional effective field theory view, from which then we switched off some fluxes, it is conceivable that doing so turns on scalars not included in our original consistent truncation. The way we are doing things, we are blind to this possibility (which will not impact scale-separated setups such as the DGKT scenario that we discuss in Section \ref{section:DGKT}), and so \eqref{eq:mass_result} should only be taken as an educated guess for further exploration. We have checked that \eqref{eq:mass_result} satisfies directly the massive IIA equations of motion in ten dimensions {provided that $c_{F_0}^2/c_R^2 = (m/(6\sqrt{30}))^2$}, and is compliant with the bound on the size of strongly coupled regions in massive IIA proposed in \cite{Aharony:2010af} (although it exactly saturates it since the inverse of curvature radius and the string dilaton decrease at the same rate as $y\rightarrow 0$), so at least there's that.

\subsection{More general examples}

We end this Section with a few comments about examples with known holographic duals that are more involved, either because they have many fields and/or non-trivial field profiles in the internal dimensions. 

Consider the AdS$_6$ vacua of Jafferis-Pufu \cite{Jafferis:2012iv} arising in Type I' string theory. The CFT comes from a system of D4-branes intersecting D8-branes and O8-planes. The bulk gravitational theory contains an AdS$_6\times S^3$ factor arising upon compactification of Type I' including $F_4$-flux. Recall that Type I' is equivalent to Type IIA on an interval, with two O8$^-$-planes at the endpoints of the interval and D8-branes to cancel the tadpole. The D8-branes induce a non-trivial value of the Romans mass in the interval. Consequently, the low-energy theory contains two geometric moduli (in addition to the dilaton) which parametrize the volume of the $S^3$ and the size of the Type I' interval. The Romans mass, the curvature of $S^3$ and the $F_4$-flux supporting the internal space will all contribute to the effective bulk potential. However, this is not all. This example exhibits a feature that did not appear in previous examples: the dilaton has a non-trivial profile in the internal dimensions, so it is not a homogeneous solution. In particular, the dilaton non-trivial profile occurs along the interval in between the O8's. This induces an additional contribution to the six-dimensional potential coming from the spatial gradient of the dilaton. In order to find the geometry probed by the D4-branes, we would need to solve the flow equations taking all these ingredients into account (except for the $F_4$-flux sourced by the D4's), which becomes quite involved as the running of the scalars will occur along both external and internal directions.

The basic issue we are facing is again that we do not have an off-shell scalar potential including all relevant fields in this case. However, there is a way to simplify the computation by performing the flux backtracking procedure in two separate steps (which secretly use the fact that we know what the final answer should be). Notice that we actually have two independent flow directions: the non-compact radial direction of the six-dimensional space and the nine-th coordinate associated to the internal Type I' interval. Hence, we can first consider the Type I' flow in nine dimensions driven by the Romans mass, and then further compactify to consider the six-dimensional flow driven by the curvature of the $S^3$ as in a Freund-Rubin setup.

Let's start with the first step and look for a running ten-dimensional solution of the form
\begin{equation}
    ds^2_{10}=dr^2+e^{2A(r)}ds^2_9,
\end{equation}
where $ds_9^2$ is a Minkowski metric.
The flow is only driven by the Romans mass since it is the only ingredient affecting the warp factor $A(r)$ and the dilaton. Using \eqref{eq:flowequations}, this leads to $P(\phi) = e^{5\phi/4}$, with $e^\phi = r^{-4/5}$ and $A(r) = \frac{1}{25} \ln r$. After going to string frame and redefining the radial coordinate $r \mapsto r^{4/5}$, 
 the result is the very well-known Polchinski-Witten solution of Type I' given by
\begin{equation}
   ds^2_{10}=dr^2+r^{-2/5}ds^2_9, \quad e^\phi \sim r^{-1}.
\end{equation}
One can also check that this is the near-horizon geometry in a background induced by D8-branes.

Next, we further compactify this nine-dimensional Minkowski space on $S^3$. The combination of the $F_4$-flux and the curvature yields the AdS$_6$ solution. However, to find the geometry probed by the D4-branes, we get rid of the $F_4$-flux and simply find the running solution driven by the curvature term.  The computation is analogous to that of Freund-Rubin compactifications in Section \ref{subsection:FR}, so we can borrow the results to obtain
\begin{aleq}
    ds_{9}^2 =  \mathcal{V}^{-1/2}(d\tilde r^2+e^{2 A(\tilde r)}ds^2_5)+\mathcal{V}^{2/3}ds^2_{S^3}=dy^2 + ds_5^2 + y^2ds_{S_3}^2,
    \label{ds9}
\end{aleq}
with $y=\tilde r^{4/7}$, $A(\tilde r) = \frac{3}{7}\ln r$ and $e^\phi$ being a constant function of $\tilde r$. The result is that the nine-dimensional metric is still flat space, so when combining everything together we get that the geometry probed by the D4's is
\begin{aleq}
    ds_{10}^2 = dr^2 + r^{-2/5} ds_9^2, \quad e^\phi \sim r^{-1},
\end{aleq}
with $ds_9^2$ corresponding to Minkowksi space as in \eqref{ds9}.

Another example that we did not discuss about is the AdS$_7$ vacua of \cite{Apruzzi:2017nck}. These vacua are notoriously complicated, involving ``football'' geometries with an anisotropic internal manifold on the bulk side, and intersecting D6-NS5-brane stacks on the other. This is yet another example  of the cases (mentioned at the end of Section \ref{s2}) where we do not have an off-shell flux superpotential. As a result, there is no guarantee that sticking to just one scalar (e.g. the volume of the compactification manifold, as we have done in previous cases) is a consistent truncation, and in fact, we have checked that if one does so, the flux backtracking procedure gives wrong results. Since we know that the internal manifold is anisotropic, it makes sense that we should include at least another modulus (and possibly more). While in principle this resolves the contradiction, it would be nice to explore multi-field cases in detail.

\section{Vacua with unknown holographic dual\label{sec:noholo}}
Now that we have some confidence in the validity of flux backtracking, we apply it to the cases of real interest: AdS vacua without a known holographic dual.

\subsection{DGKT scale-separated AdS\texorpdfstring{$_4$}{4} vacua in massive IIA}\label{section:DGKT}
The DGKT \cite{DeWolfe:2005uu, Camara:2005dc} vacua are compactifications of massive IIA string theory on a Calabi-Yau manifold in the presence of unbounded $F_4$-fluxes, bounded $H_3$-flux (and $F_0$-flux) and O6-planes. This is an example of special interest as it provides a scale-separated AdS$_4$ solution, namely the size of the extra dimensions is smaller than the AdS size, so the vacuum is honestly four-dimensional. However, it is not clear whether this proposed supergravity solution can be uplifted to a UV-consistent string theory solution (see \cite{Coudarchet:2023mfs} for a review on potential issues). Recently, we have found \cite{Montero:2024qtz} that this vacuum is in tension with the vanilla application of the Weak Gravity conjecture to domain walls, so either there is some inconsistency in the DGKT vacuum or the Weak Gravity Conjecture for domain walls must be reformulated, which would also have profound implications for the expected instability of non-SUSY vacua \cite{Ooguri:2016pdq}. Any progress in constructing the CFT dual of this AdS solution could help to settle the debate and find out whether the vacuum is UV consistent or not. Notice that, at the moment, there is no single example of an AdS/CFT pair exhibiting scale separation; all AdS proposed scale-separated vacua do not have known CFT duals yet, and whether this is at all possible has been put into question \cite{Polchinski:2009ch,Perlmutter:2024noo,Lust:2019zwm,Collins:2022nux,Montero:2022ghl}. Therefore, DGKT vacua (or similar constructions) have the potential to become the first scale-separated examples in the literature, if a CFT dual can be found. 

\begin{figure}[htb!]
\begin{center}
\includegraphics[scale = 0.3]{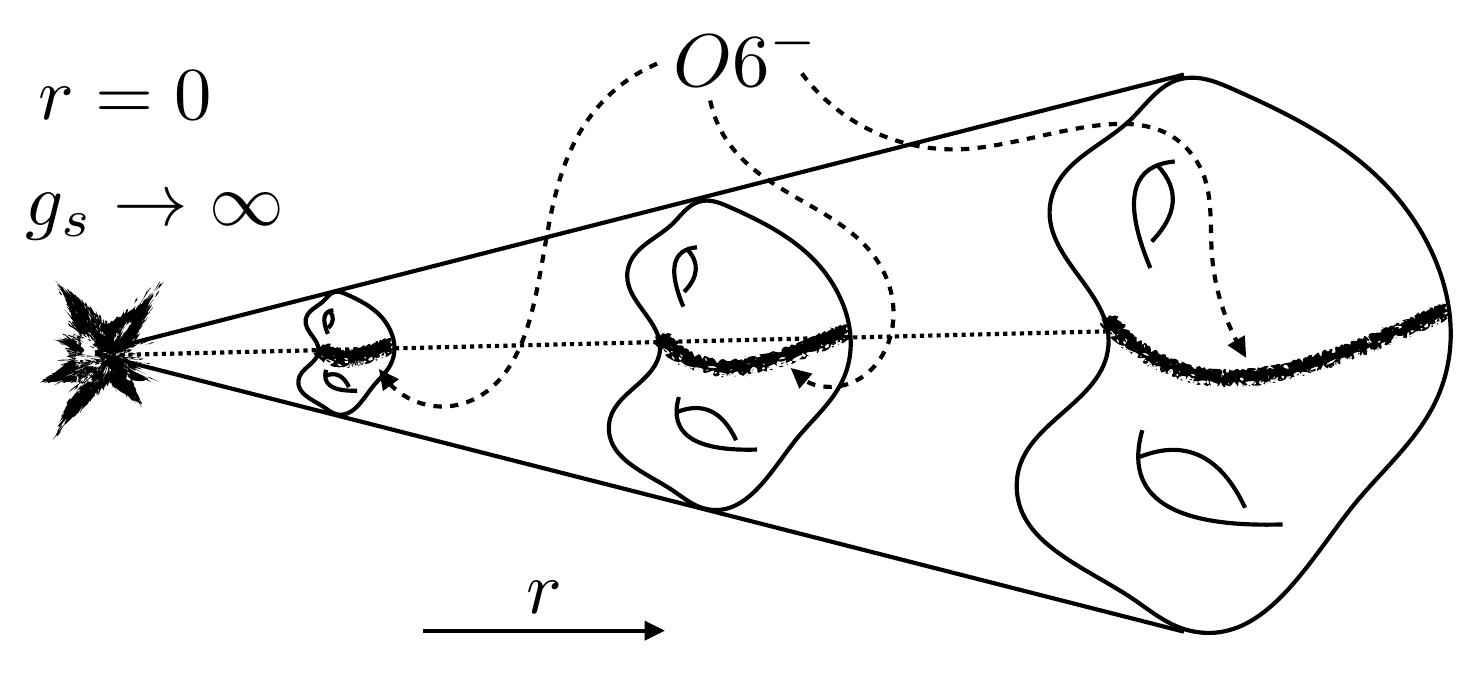}
\caption{Flux backtracking applied to DGKT produces a strongly coupled singularity in massive IIA, depicted in the figure. The $CY_3$ of DGKT is fibred along the $r$ direction, reaching 0 at finite distance. Probing this singularity with D4-branes and taking the near-horizon limit should return the original DGKT vacuum. Notice that the $O6^-$ plane (depicted as a fuzzy line on each of the three illustrative slices) of DGKT ends at the singularity; this is possible because the geometry is also threaded with $H_3$-flux, cancelling the tadpole.}
\label{moodeng}
\end{center}
\end{figure}

A first step towards constructing the CFT dual is to find the brane picture of the AdS vacuum, which we will do in this section. The $F_4$-flux number $N$ parametrizes the amount of control and scale separation, since $g_s^{-1}$, the overall volume and the ratio between the KK and the AdS energy scales grow with $N$. Hence, we will apply backtracking to this flux only, so the result will be some singularity in massive IIA where an $O6$-plane and $H_3$-flux end, as depicted in Figure \ref{moodeng}. Probing this singularity with D4-branes should reproduce the AdS DGKT background as their near-horizon geometry. See  \cite{Kounnas:2007dd} for previous attempts involving smeared sources and a toroidal cover of the Calabi-Yau manifold. 

When switching off the $F_4$-flux, we can study the problem using the explicit 4d $\mathcal{N}=1$ solution of \cite{DeWolfe:2005uu}. Since the solution is scale-separated, the effective field theory captures in principle the full dynamics of all scalars. The remaining fluxes then generate the following potential
\textcolor{black}{
\begin{equation}\label{DGKT}
    V = \dfrac{1}{s^3}\left[\dfrac{A_{F_0} u^3}{s}+\dfrac{A_{H_3} s}{u^3}-A_{O6}\right],
\end{equation}
}
for which the potential $P$ is
\begin{aleq}
    P = c_{F_0} u^{\frac{3}{2}}s^{-2} + c_{H_3} u^{-\frac{3}{2}}s^{-1} 
\end{aleq}
and the flow equations \eqref{eq:flowequations} are solved by the following running solution
\begin{aleq}\label{DGKTsol}
    s(r) \sim r^{2/3}, \quad u(r) \sim r^{2/9}, \quad A(r) = \frac{13}{27} \ln r.
\end{aleq}
The geometry that the D4-branes should probe in the DGKT geometry is then an orientifold of 
\begin{aleq}\label{eq:dgktsing}
    ds_{10}^2 = dy^2 + y^{-10/9} ds_3^2 + y^{2/3} ds_{CY}^2, \quad g_s(y) \sim y^{-1}
\end{aleq}
where $y= r^{1/3}$.  

Result \eq{eq:dgktsing} is, to our knowledge, new; it unambiguously comes out from the flux backtracking procedure applied to DGKT. 
Unlike previous examples, this resulting singularity is neither conical nor deformed conical. This is to be expected, since massive IIA does not admit flat space as a solution; asymptotically conical geometries precisely asymptote to flat space. The interesting behavior however is the singularity at $y\rightarrow0$. We observe that the string coupling diverges there. This is even more interesting, since the fact that the coupling becomes strong means that we will never be able to understand this singularity via D-brane techniques. In other examples (such as ABJM, discussed in Section \ref{ABJM}), the reason why we understand the system is that there is a UV brane picture involving weakly coupled singularities, which can be probed via D-branes in string perturbation theory. Once the UV dynamics is known, it may flow to a strongly coupled system in the IR (like for ABJM for small $k$), but the QFT description provides a foothold from which we can understand the system explicitly, providing e.g. an upper bound for the central charge via the c-theorem. Since the DGKT singularity is strongly coupled, we cannot do any of the above, even in principle (see e.g. \cite{Banks:2006hg} for a proposal of increasing the number of degrees of freedom at strong coupling by considering multi-prong strings).
We want to note, though, that the central charge is much larger than the $N^{5/3}$ scaling obtained in the massive IIA solution of \cite{Guarino:2015jca}, for which the scaling of the central charge is smaller than $N^2$ even though it is also strongly coupled.

Finally, reference \cite{Aharony:2010af} gave evidence that massive IIA cannot be strongly coupled. The precise bound they put forth was that the string coupling could not grow beyond the local curvature scale in string units, although most of the evidence from this statement was obtained in the supergravity approximation.
Hence, the bound that needs to be satisfied is $g_s\lesssim l_s/R$ where $R$ is the radius of curvature. Using \eqref{DGKTsol} and \eqref{us}, we obtain that $g_s\sim y^{-1}\lesssim l_s/R\sim y^{-1/3}$, where we have also used that the curvature radius in string units goes as $\mathcal{V}^{1/6}\sim u^{1/2}$. Hence, this bound holds as long as we are at a distance $y$ from the singularity bigger than a string length (at smaller distances we cannot trust the solution anyway), so the solution is consistent with \cite{Aharony:2010af}.

\subsection{Scale-separated AdS\texorpdfstring{$_4$}{4} vacua in massless IIA/M-theory}

The DGKT vacuum is not the only proposed scale-separated AdS vacuum in the literature.
For instance, by performing two T-duality transformations on the DGKT solutions and doing a rescaling of fluxes, new scale-separated AdS$_4$ solutions were obtained in IIA without Romans mass \cite{Cribiori:2021djm}. These involve an unbounded $F_6$-flux, partly unbounded $F_2$-flux that generates the scale separation, and curvature of the internal Iwasawa manifold. The $F_2$-fluxes are distributed anisotropically over the internal manifold, resulting in an anisotropic scale separation. This setting has the advantage of allowing for strongly coupled solutions and hence an uplift to M-theory due to the absence of Romans mass. The backreaction of the O-planes is better understood in this context \cite{Cribiori:2021djm}. From an M-theory perspective, since all fluxes but $F_6$ are geometrized, this solution is of Freund-Rubin type, and should arise from compactifying M-theory on a 7-dimensional manifold that admits a weak $G_2$ structure \cite{Cribiori:2021djm}; equivalently, this means that the 7-manifold equipped with the scale-separated metric is the base of an eight-dimensional $Spin(7)$ cone. However, it is not clear that such a $G_2$ structure exists. Progress towards these questions can be found in \cite{VanHemelryck:2024bas}.

In the following, we derive the singular geometry that should be probed by D2-branes in order to reproduce this AdS vacuum. When removing the $F_6$-flux and replacing it with D2-branes, the remaining core ingredients that generate the flow are the same as in the ABJM construction (Section \ref{ABJM}): $F_2$-flux and curvature. While details and the precise flux configuration may differ, the overall dependence of the universal moduli on the radial coordinate will remain the same, yielding
\begin{aleq}
    ds_{10}^2 = dy^2 + y^{2/3}ds_3^2 + y^2ds_{\text{Iwasawa}}^2,\quad g_s(y) \sim y
\end{aleq}
in ten dimensions, and 
\begin{aleq}\label{eq:11dmetric}
    ds_{11}^2 = d\Tilde{y}^2 + ds_3^3 + \Tilde{y}^2 ds^2_{\text{Iwasawa}\times S^1},
\end{aleq}
in eleven dimensions, where $\tilde{y} = y^{2/3}$. To derive this result, we have borrowed the result for the running solution in \eqref{abjm1} and \eqref{abjm3}, but taking into account that the internal manifold is now the Iwasawa manifold. Here, the unbounded $F_2$-flux has been absorbed into the internal metric component. More details and the explicit solution including the flux dependence of each metric factor can be found in Appendix \ref{appendix:details}.

Interestingly, the resulting metric is quite simple and corresponds to a conical singularity, unlike in the above DGKT setup; this is also what one would expect from the $G_2$ cone picture described above. Moreover, the singularity is weakly coupled as $g_s$ vanishes as $y\rightarrow 0$, so one could aim to understand the growth of the central charge using worldsheet techniques. According to the bulk AdS solution, the central charge of the dual CFT should grow as $c\sim l_{AdS}^2M_{p,4}^2\sim N^{3/2}k_1^3 k_2^{-5/2}$, where $N$ is the $F_6$-flux,  dual to the D2-branes probing the weakly coupled singularity, and $k_1$, $k_2$ are different components of $F_2$-flux that lift to metric fluxes in M-theory. This is consistent with the expected $N^{3/2}$ scaling of M2-branes probing a conical singularity. The extra  $k^3_1k_2^{-5/2}$ factor should then arise from the particular singular geometry in which the D2-branes are placed. The flux $k_1$ is unbounded, while the component $k_2$ (corresponding to $e_{16}$ in the notation of \cite{Cribiori:2021djm}) is fixed by the D6-brane tadpole. We keep it here to illustrate that for $k_1\sim k_2\sim k$ (which is an $O(1)$ quantity fixed by the tadpole), the above formula for the central charge is reminiscent of the small $k/N$ ABJM result $c\sim N^{3/2}k^{1/2}$, including the leading $k$-dependence. In fact, if we restore the flux dependence on the result for the metric \eqref{eq:11dmetric} (see Appendix \ref{appendix:details}), we get
\begin{aleq}
    ds_{11}^2 =
 d\Tilde{y}^2 + k_1^{4/3}k_2^{-10/9} ds_3^2 + (k_1^{-1} k_2)^{4/3} \Tilde{y}^2 ds_{Iwasawa}^2 + k_2^2 k_1^{-4} \Tilde{y}^2 dz^2
\label{IIAsing}
\end{aleq}
with $    g_s(y) = k_2 k_1^{-2} y$,
which for $k_1=k_2$ also resembles the $k$-dependence of the metric in ABJM\footnote{In ABJM, we have \begin{aleq}
    ds_{10}^2 =dy^2 + k^{-4/9}y^{2/3}ds_3^3 + y^2 ds_{\mathbb{C}P^3}^2,
\end{aleq}
with $y = k^{-1/3}r^{1/3}$, where we have restored the flux dependence in \eqref{abjm1}, \eqref{abjm2}. Using that the string coupling runs as $    g_s = e^\phi  = k^{-1} y$ and defining $\Tilde{y} = k^{1/3}y^{2/3}$,
the metric gets uplifted to 11d as follows,
\begin{aleq}
    ds_{11}^2 &= e^{-2\phi/3}ds_{10}^2 + e^{4\phi/3} dz^2= d\Tilde{y}^2 + k^{2/9} ds_3^2+ \Tilde{y}^2 (ds_{\mathbb{C}P^3}^2+k^{-2}dz^2).
\end{aleq}}. We may understand this as arising from the fact that in both cases we have a IIA picture where the $F_2$-flux scales homogeneously with a single integer $k$.

 The main difference with ABJM gets manifest  when $k_1$ is taken parametrically large and independent of $k_2$, so that $l_{AdS}/l_{KK}\sim k_1^{1/2}$ and the solution becomes scale-separated. If $N\sim k_1$ then we also recover the same huge scaling of the central charge as in DGKT $c\sim N^{9/2}$. Hence, in any parametrically scale-separated solution within this family of $AdS_4$ vacua, the central charge must grow strictly faster than $N^{3/2}$.

 In ABJM, when $k\sim N$, the supergravity solution breaks down since the t'Hooft coupling is given by $\lambda=l^4_{c}/l_s^4=N/k$, so the internal curvature $l_c$ becomes of string size. However, in this other family of $AdS_4$ vacua, the internal volume in string units scale as $N^{3/2}k_1^{-1}k_2^{-1/2}$ with $k_2$ fixed by the tadpole, so that the supergravity regime remains valid as long as $k_1\ll N^{3/2}$. Still, due to the similarities with ABJM, one could be tempted to imagine that the dual theory is described in the UV by a gauge theory (associated to the M2-branes probing the corresponding conical singularity). However, we do not know how to make this expectation consistent with the supposed growth of the central charge. For a gauge theory with a simple gauge factor, the central charge should grow as $N^2$ for fixed t'Hooft coupling\footnote{This is indeed satisfied in ABJM, where $c\sim N^{3/2}k^{1/2}\sim \frac{N^2}{\sqrt{\lambda}}$ for $\lambda=\frac{N}{k}\gg 1$.}. If we assume the same dependence as in ABJM, where $\lambda=l^4_{c}/l_s^4$ with $l_c$ being the curvature of the internal space in string units, then we would get $c\sim \frac{N^6}{\lambda^{9/4}}k_2^{-5/2}$ for these scale-separated AdS vacua. If additionally we set $k_2\sim N/\lambda_2$, the central charge is of the form $c\sim N^2 f(\lambda,\lambda_2)$. This suggests that the dual CFT (if it exists) could flow from a gauge theory (whose central charge scales with $N^2$ times a function of 't Hooft couplings) only if $k_2$, which is fixed by the tadpole, is also related to some t'Hooft coupling of a gauge factor that somehow remains always strongly coupled. It would be interesting to further explore this in the future.
 

\subsection{Scale-separated AdS\texorpdfstring{$_3$}{3} vacua in massive IIA}

There have also been plenty of similar proposals of three dimensional --rather than four dimensional-- scale-separated vacua in massive Type IIA using G2-holonomy manifolds instead of Calabi-Yau's.
The construction presented in \cite{Farakos:2020phe} is very similar to the DGKT vacua with similar ingredients such as an unbounded $F_4$-flux, Romans mass and an $H_3$-flux. In this way, they obtain scale-separated AdS$_3$-vacua, where the $F_4$-flux again parametrizes the amount of scale separation, and the internal manifold is instead a G2-holonomy manifold.
\textcolor{black}{
After removing the unbounded $F_4$-flux from the action, the residual scalar potential and superpotential take the form
\begin{equation}\label{V3dIIA}
   V = A_{F_0}u^{\frac{7}{2}}s^{-3}+ A_{H_3}u^{-3} s^{-2}-A_{O6}u^{\frac{1}{4}}s^{-\frac{5}{2}}, \quad P = c_{H_3} u^{-\frac{3}{2}}s^{-1} + c_{F_0} u^{\frac{7}{4}}s^{-\frac{3}{2}},
\end{equation}
}
The flow generated by the Romans mass and the NSNS-flux in three dimensions is solved by
\begin{aleq}
    s(r) \sim r^{13/16}, \quad u(r) \sim r^{1/8}, \quad A(r) = \frac{11}{16} \ln r,
\end{aleq}
which results in the following ten-dimensional metric
\begin{aleq}
    ds_{10}^2 &= [s(r)]^{-2} \left(dr^2 + e^{2A(r)}ds_2^2\right) + [u(r)]ds_{G_2}^2
    &= dy^2 + y^{-4/3}ds_2^2 + y^{2/3}ds_{G_2}^2,
\end{aleq}
with $y=r^{3/16}$ and $g_s(y) = y^{-1}$. This is the singularity that should be probed by domain walls consisting of D4-branes wrapped on 3-cycles in order to reproduce the above AdS$_3$ geometry as its near-horizon limit. This geometry turns out to be very similar to the geometry found for the DGKT vacua \eqref{eq:dgktsing} and is also strongly coupled. Again, we have a very large central charge $c\sim N^4$ \cite{Apers:2022zjx}, but we cannot say much in this regard since the singularity is strongly coupled. We have also checked that the flow is consistent with the criterium of \cite{Aharony:2010af} of not having strongly coupled supergravity configurations of masisve IIA, since here the curvature blows up faster than $g_s^{-1}$ as $y\rightarrow 0$.

\subsection{KKLT AdS\texorpdfstring{$_4$}{4} vacua}

Another important proposed AdS vacuum whose CFT dual is unknown is the 4d $\mathcal{N}=1$ AdS vacuum in the KKLT scenario. This vacuum serves as the starting point before uplifting to a de Sitter solution, and recent advances have provided concrete progress in its construction within string theory \cite{Demirtas:2021nlu, Demirtas:2021ote, McAllister:2024lnt}. Despite this progress, it remains unclear whether the vacuum is fully protected from all possible corrections, as it contains blow-up cycles of string size and the very little amount of supersymmetry makes its control very challenging. Unlike DGKT, this vacuum cannot exist at parametrically large values of the volume, and exhibits an even larger scale separation, as the internal space grows logarithmically with the AdS scale.\footnote{In particular, one has $l_{KK} \sim \sigma$ and $l_{AdS} \sim \sqrt{\sigma} e^{a \sigma}$ with $\sigma$ parametrizing the overall volume.} This seems to imply that the dual CFT, if it exists, should have a central charge that grows exponentially on $N$ rather than polynomially (assuming that $N$ is still related to the flux quanta).

The papers  \cite{Lust:2022lfc, Bena:2024are} precisely attempted to construct brane picture of this vacuum, by dualizing all fluxes to branes. 
When these branes are placed in flat space, neglecting non-perturbative corrections, \cite{Lust:2022lfc, Bena:2024are}  show that their worldvolume degrees of freedom are insufficient to explain the very large growth of the central charge of the putative dual CFT. While these branes in flat space alone are not expected to provide the CFT dual of KKLT,  those papers aim to argue that the resulting worldvolume theory is a UV description that should flow to the actual CFT dual to KKLT in the IR, and therefore provide an upper bound on the central charge. Rather than entering into the discussion of the consistency, or even existence, of this putative RG flow, we are interested here in whether it is possible to get information directly about the singularity that should be probed by the branes in order to realized the putative CFT dual to KKLT. If a brane picture for KKLT exists, the branes will most likely have to probe some singular, complicated geometry in order to produce the KKLT AdS vacuum as their near-horizon geometry. A natural question is, therefore, whether we can apply our flux backtracking procedure to determine such singular geometry where the branes should be placed to yield the dual field theory of KKLT. 

Let us first describe the low-energy effective theory for KKLT. The idea is to first introduce $G_3$ fluxes that stabilize the dilaton and all complex structure moduli of the Calabi-Yau threefold. The overall volume of the Calabi-Yau is then stabilized by considering non-perturbative corrections (e.g. from euclidean D3-branes), such that the superpotential has the following structure:
\begin{equation}
\label{WKKLT}
    W=W_0+ce^{-2\pi a T},
\end{equation}
where $W_0$ includes the flux dependence and it is evaluated at the flux stabilized values of the complex structure moduli and dilaton.
The low-energy scalar potential for the K\"ahler modulus $T$ becomes 
\begin{equation}
    V=\frac{\pi a c e^{-2\pi a c}}{\sigma^2}\left(\frac{2\pi a c \sigma e^{-2\pi a c}}3+W_0+c e^{-2\pi a c}\right)
    \label{VKKLT}
\end{equation}
where $\sigma=Re(T)$ parametrizes the overall volume of the CY.

If we try to apply the flux backtracking method to the solution, we quickly see that there are key differences between KKLT and the other setups studied in this paper. 
The main difference is that there is no flux that can be taken parametrically large while keeping a supersymmetric solution, since all involved fluxes are bounded by the D3-tadpole condition. Hence, the first step of the flux backtracking procedure may already be more difficult, since finite N corrections cannot be scaled away. Nevertheless, we may still try to switch off some flux, even if it enters in the tadpole, dualize it to a brane and see where this takes us.

When switching off a flux, the tadpole forces us to include additional D-branes in the
compactification, which will change the low-energy effective field theory. In a generic position, each of the branes will contribute an open string sector with an $N=4$ U(1) vector multiplet. At weak coupling, we can decouple the dynamics of these open string fields from the closed string modes, so that we can study the running solution considering only the latter. Moreover, the D3-brane tension will can cancel against the orientifold, as happens for the ISD fluxes in KKLT. This seems to suggest that we can still use \eqref{WKKLT} as a good approximation for the superpotential of the running solution. However, we should keep in mind that this approximation might fail close to the singularity where small volume or strong coupling effects can imply that open and closed string sectors do no longer decouple.
Modulo this caveat, we can now check what the result would be for the singular geometry depending on which fluxes are dualized to branes. 

In Appendix \ref{app:KKLT}, we study two particular cases: When all fluxes are removed (so $W_0=0$ classically), and the case where only $F_3$-flux is removed, so that the stack of branes would consist only of D5-branes. When all fluxes are removed, the remaining non-perturbative superpotential generates a running solution towards small overall volume and strong coupling. At a first glance, it might be surprising that we obtain the same qualitative behaviour as in a Freund-Rubin compactification, given the very different nature of the flux vacuum: when removing a single unit of flux in the AdS vacuum, Freund-Rubin goes towards smaller values of the volume while KKLT does the opposite. In Appendix \ref{appendix:details} we show that this is a result of a friction term in the equations of motion driving the running solution towards small volumes in both cases. Finally, if only $F_3$-flux is removed, we get similar behaviour if the remaining NS flux and the non-pertubative terms contribute with the same sign to the superpotential.

We wish to end this Section with a note of caution. The flux backtracking procedure only yields the crudest features of the would-be singularity, and cannot be used to conclusively establish its existence or properties. In other words, while we found no obstruction to the existence of a KKLT singularity with the right asymptotics, there might be other issues invisible to our analysis, which may however be taken as a first step towards a KKLT brane picture.

\section{Conclusions}\label{s5}
In this brief note we have explored the application of a ``flux backtracking'' procedure, allowing one to reverse-engineer a brane picture starting from a given AdS flux compactification. The main tool required is an off-shell scalar Lagrangian for the low-dimensional theory. Using this, one can construct a dynamical cobordism (in the sense of \cite{Buratti:2021fiv, Buratti:2021yia, Angius:2022aeq}) leading to the desired geometry. This Lagrangian could correspond either to a consistent truncation of the full system or to the low-energy effective field theory if the AdS vacuum is scale-separated.

We have checked this procedure in a number of known AdS/CFT pairs. In all these examples, the flux backtracking procedure reproduces the correct result for the possibly-singular geometry that should be probed by the stack of branes in order to generate the corresponding AdS background as its near-horizon geometry. More interestingly, we have applied it also to AdS flux vacua without known holographic dual, such as \cite{Guarino:2015jca} or the scale-separated DGKT vacuum of \cite{DeWolfe:2005uu}. Doing this, we have produced a candidate massive IIA singularity which we believe that, when probed with D4-branes, results in the DGKT solution as a near-horizon geometry. The question of whether this vacuum is UV consistent and has a CFT dual reduces then to whether the worldvolume theory of the D4-branes in this singular geometry flows to a suitable CFT. The singularity is strongly coupled and preserves 3d $\mathcal{N}=1$ supersymmetry only, which is not protected against quantum effects, so the system will be difficult to analyze explicitly. At the singularity, curvatures blow up; it would be interesting to elucidate its fate given the low supersymmetry of the system, but at present we do not know how to do this. However, we hope that our explicit derivation of the brane picture could help in the future towards either finding the CFT dual or showing that it does not exist.

We have also studied the scale-separated solution of \cite{Cribiori:2021djm}, which is a massless IIA ``cousin'' of the DGKT solution. From the point of view of our procedure, this case is completely analogous to ABJM, and the resulting singularity is compatible with the natural picture that the system is obtained via a stack of M2-branes probing a $Spin(7)$ conical singularity in M-theory. However, whether an appropriate family of singularities exists is left for future work (see also \cite{VanHemelryck:2024bas}).

Finally, it is natural to try to flux backtrack KKLT vacua, particularly given the recent concrete progress and discussions on the topic \cite{Demirtas:2021nlu, Demirtas:2021ote, McAllister:2024lnt,Lust:2022lfc, Bena:2024are}.  A key difference with the previous setups is that there is no unbounded flux, so the procedure is sensitive to finite $N$ corrections. Ignoring this, we have found several singularities, which asymptote to Ricci flat geometries, and which if consistent (a question about which we have nothing to say) and if probed by appropriate stacks of branes, would yield the CFT dual to KKLT. In all cases studied, however, the singularity is either non-perturbative or must be probed by non-perturbative stacks of branes, which difficults using this perspective to learn something new about KKLT.

More generally, flux backtracking or other similar procedures can be applied to all sorts of vacua where the brane picture is not known, but nevertheless one has some control over the low-energy effective bulk action. Hopefully, doing so will shed some light on these questions and illuminate new corners of the AdS/CFT Landscape.

\vspace{5mm}
\textbf{Acknowledgements}: We are indebted to Nikolay Bobev, Joe Conlon, Marco Fazzi, Suvendu Giri, and Thomas Van Riet and  for enlightening discussions.  The authors thank Iberian Strings '23 and the Simons Center for Geometry and Physics for hospitality during the Summer Workshop '23. We also thank the Erwin Schrodinger International Institute for Mathematics and Physics for their hospitality during the programme ``The Landscape vs. the Swampland'', and the Aspen Center for Physics, which is supported by National Science Foundation grant PHY-2210452, during the program ``Bootstrap, Holography and Swampland''for hospitality during the final stages of this work. This work was partially supported by a grant from the Simons Foundation. MM also thanks CERN for hospitality during the completion of this work. MM is currently supported by the RyC grant RYC2022-037545-I and project EUR2024-153547 from the AEI and was supported by an Atraccion del Talento Fellowship 2022-
T1/TIC-23956 from Comunidad de Madrid in the early stages of this project. M.M. and I.V. thank the KITP program ``What is String Theory?" for providing a stimulating enviroment for discussion. The authors thank the Spanish Research Agency (Agencia Estatal de Investigacion)
through the grants IFT Centro de Excelencia Severo
Ochoa CEX2020-001007-S and PID2021-123017NB-I00,
funded by MCIN/AEI/10.13039/501100011033 and by
ERDF A way of making Europe. 
The work of I.V. was supported by the grant RYC2019-028512-I from the MCI (Spain), the ERC
Starting Grant QGuide101042568 - StG 2021, and the Project ATR2023-145703 funded by MCIN /AEI /10.13039/501100011033. FA is supported by the Clarendon Scholarship in partnership with the Scatcherd European Scholarship, Saven European Scholarship, and the Hertford College Peter Howard Scholarship.

\appendix
\section{Dimensional reduction of the ten-dimensional Type IIA action with fluxes\label{app:IIA}}
The ten-dimensional bosonic Type IIA action is of the form
\begin{aleq}
    S_{10} &= \frac{1}{2\kappa_{10}^2} \int d^{10}x \, \sqrt{-g} \, e^{-2\phi} \left( R + 4(\partial_\mu \phi)^2 - \frac{1}{2} |H_3|^2 - e^{2\phi} \sum_p |F_p|^2 \right)\\
    & + \text{local contributions},
\end{aleq}
where local contributions arise from D-branes and/or O-planes. Consider a compactification of this theory to $d$ dimensions.
We are interested in determining the dependence of the $d$-dimensional potential on the universal moduli $u$ and $s$. The internal volume in string frame is
\begin{aleq}
    \mathcal{V} = u^{\frac{10-d}{2}},
\end{aleq}
and the ten-dimensional dilaton $\phi$ is related to these moduli by
\begin{aleq}
    e^{-2\phi} = s^{d-2} u^{\frac{d-10}{2}}.
\end{aleq}
The canonically normalised scalars in terms of these universal moduli are
\begin{aleq}
    \overline{u} \equiv \sqrt{\frac{10-d}{4}} \ln u, \quad \overline{s} \equiv \sqrt{d-2} \ln s.
\end{aleq}
The moduli dependence of the $d$-dimensional scalar potential is as follows
\begin{aleq}
    V &= A_R u^{-1} s^{-2} + A_{H_3} u^{-3} s^{-2} + \sum A_p u^{\frac{10-d-2p}{2}} s^{-d}\\
    &+\text{local contributions},
    \label{VIIA}
\end{aleq}
where the $A$'s are coefficients that we do not specify.
The Ansatz `superpotential' to reproduce this potential is
\begin{aleq}
\label{PIIA}
    P = c_R u^{-\frac{1}{2}} s^{-1} + c_{H_3} u^{-\frac{3}{2}} s^{-1} + \sum c_p u^{\frac{10-d-2p}{4}}s^{-\frac{d}{2}},
\end{aleq}
where we again do not specify constants as we only need the moduli dependence. The local sources do not contribute to the superpotential.

\section{More details on scale-separated AdS\texorpdfstring{$_4$}{4} vacua in IIA with M-theory uplift}\label{appendix:details}
In this appendix, we provide additional details on the computation of the flow in the scale-separated  AdS$_4$ vacua in massless IIA. In particular, we will give the flux dependence, illustrating the potentially non-isotropic scalings in this case.
The EFT is described by the following schematic 4d $\mathcal{N}=1$ superpotential (ignoring axionic contributions) and K\"ahler potential,
\begin{aleq}
    W = c_{F_6}-c_{c}u_1 s + c_{F_{2, 1}}u_1 u_2  + c_{F_{2, 2}} u_2^2, \quad K = - \ln s^4u_1 u_2^2,
\end{aleq}
with contributions from unbounded $F_6$-flux, curvature and both unbounded and bounded $F_2$-flux. For the case of a toroidal internal manifold, $u_1$ parametrises the size of the first sub-torus, while $u_2$ corresponds to the size of the remaining two sub-tori. We label the flux quanta as follows,
\begin{aleq}
    c_{F_6} \sim N, \quad c_{F_{2, 1}}\sim k_1, \quad c_{F_{2,2}} \sim k_2,
\end{aleq}
where both $N$ and $k_1$ can in principle be arbitrarily large.
The central charge scales with these fluxes as
\begin{aleq}
    c \sim N^{3/2} k_1^3 k_2^{-5/2}.
\end{aleq}
The unbounded $F_2$-flux is the one responsible for scale separation, as
\begin{aleq}
    \dfrac{L_{AdS}}{L_{KK}} \sim k_1^{1/2}.
\end{aleq}
We note that for $k_1 \sim N \gg 1$, we reproduce the DGKT scaling relations, while for $k_1 \sim k_2 \sim 1$, we obtain the ABJM scaling relations.\\
To find the geometry probed by D2-branes dual to the unbounded $F_6$-flux, we consider the flow generated by the following residual superpotential
\begin{aleq}
    W = -c_{c}u_1 s + c_{F_{2, 1}}u_1 u_2  + c_{F_{2, 2}} u_2^2, \quad K = - \ln s^4u_1 u_2^2,
\end{aleq}
The flow equations are
\begin{aleq}
    &s' = \frac{s \left(2 k_2 u_2^2+u_1 \left(s-2 k_1 u_2\right)\right)}{\sqrt{s^4 u_1 u_2^2}}, \quad u_1' = \frac{2 u_1 \left(k_2 u_2^2+u_1 \left(k_1 u_2-s\right)\right)}{\sqrt{s^4 u_1 u_2^2}}\\
    & u_2' = \frac{2 \left(s u_1 u_2-k_2 u_2^3\right)}{\sqrt{s^4 u_1 u_2^2}}, \quad A' =-\frac{u_1 \left(k_1 u_2-s\right)-k_2 u_2^2}{\sqrt{s^4 u_1 u_2^2}},
\end{aleq}
and a solution is given by
\begin{aleq}
    &s(r) = \frac{3}{2} \sqrt[3]{\frac{3}{5}} \sqrt[3]{k_2} r^{2/3} \sim k_2^{1/3} r^{2/3},\\
   &u_1(r) = \frac{27 \sqrt[3]{\frac{3}{5}} k_2^{4/3} r^{2/3}}{20 k_1^2} \sim k_1^{-2} k_2^{4/3}r^{2/3}, \\
   & u_2(r) = \frac{9 \sqrt[3]{\frac{3}{5}} \sqrt[3]{k_2} r^{2/3}}{8 k_1}\sim k_1^{-1} k_2^{1/3} r^{2/3},\\
    & A(r) = \dfrac{7}{9} \ln r.
\end{aleq}
The resulting geometry is 
\begin{aleq}
    ds_{10}^2 &= (k_2^{-2/3}r^{-4/3})[dr^2 +r^{14/9} ds_3^2] + k_2^{2/3}k_1^{-4/3} r^{2/3} ds_{Iwasawa}^2\\
    &= dy^2 + k_2^{-4/9} y^{2/3} ds_3^2 + k_2^{4/3} k_1^{-4/3} y^2 ds_{Iwasawa}^2,
\end{aleq}
where $y = k_2^{-1/3} r^{1/3}$. We stress that $k_2 \sim 1$ is bounded, while $k_1 \sim M$ is unbounded and we need $M \rightarrow \infty$ to achieve scale separation,
\begin{aleq}
    ds_{10}^2 =dy^2 + y^{2/3}  ds_3^2 +M^{-4/3} y^2 ds_{Iwasawa}^2.
\end{aleq}
For the string coupling, we find
\begin{aleq}
    g_s(y) = k_2 k_1^{-2} y,
\end{aleq}
so that the uplifted metric is of the form
\begin{aleq}
    ds_{11}^2 &= d\Tilde{y}^2 + k_1^{4/3}k_2^{-10/9} ds_3^2 + (k_1^{-1} k_2)^{4/3} \Tilde{y}^2 ds_{Iwasawa}^2 + k_2^2 k_1^{-4} \Tilde{y}^2 dz^2\\
    &= d\Tilde{y}^2 + M^{4/3}ds_3^2+ M^{-4/3}\Tilde{y}^2 (ds_{Iwasawa}^2+ M^{-8/3} dz^2).
\end{aleq}
with $\Tilde{y} = k_2^{-1/3}k_1^{2/3}y^{2/3}$, which means that the uplifted M2-branes probe a conical singularity.

\section{KKLT vacua}\label{app:KKLT}
\subsection{Removing all fluxes}\label{app:KKLT1}
In KKLT, we have both fluxes and non-perturbative effects. The 3-form fluxes can be turned into D5- and NS5-branes, after which we only remain with non-perturbative effects. We will study what geometry these non-perturbative effects induce with our backtracking method. However, we need to be cautious as the fluxes we now extract are all bounded by tadpoles contrary to the previous examples we discussed. 

We turn off all fluxes in the superpotential and consider
\begin{aleq}
W = A_0 e^{-aT}, 
\end{aleq}
where $t = \text{Re}\ T$ is the volume modulus related to the total internal volume in string frame as $\mathcal{V}\sim t^{3/2}$. We consider a K\"ahler potential of the form
\begin{aleq}
    K = -\ln (T+\overline{T})^3(S+\overline{S})(Z+\overline{Z})^p,
\end{aleq}
where $S$ is the axio-dilaton and $Z$ are the complex structure moduli, and $p>0$ is a positive integer. The K\"ahler potential should take such a polynomial form in asymptotic regions of moduli space.
The flow equations (with $s = \text{Re}\  S, z = \text{Re}\  Z$) for the moduli are given by
\begin{aleq}
& t' =\frac{2 A_0 e^{-a t } (2 a t +3) z ^{-p/2}}{3 \sqrt{t  s }}, \quad
& s' = 2 A_0 e^{-a t } z ^{-p/2} \sqrt{\frac{s }{t ^3}},\\
& z' = \frac{2 A_0 e^{-a t } z ^{1-\frac{p}{2}}}{\sqrt{t ^3 s }}, \quad
    & A' =\frac{A_0 e^{-a t } z ^{-p/2}}{\sqrt{t ^3 s }}.
\end{aleq}

Solving these equations in terms of the radial coordinate \( r \) leads to complicated expressions, but we can instead solve them in terms of the metric factor \( A \),
\begin{aleq}
    t(A) = \frac{3 c_1 e^{2 A}}{1-2 a c_1 e^{2 A}}, \quad s(A) = c_2 e^{2A}, \quad z(A) =c_3 e^{2A},
\end{aleq}
for constants $c_1, c_2, c_3$. 
As we approach the singularity \( e^A \to 0 \), the volume shrinks, and we flow towards strong coupling and small complex structure.

The residual potential is strictly positive and takes the form:
\begin{aleq}
V(t, s, z) = \frac{A_0^2 e^{-2 a t } z ^{-p} \left(4 a^2 t ^2+12 a t +3 p+3\right)}{3 t ^3 s }.
\end{aleq}
Hence, the potential increases monotonically as the moduli decrease. The flow towards the singularity is therefore an uphill flow. This contrasts with all previous pure-flux examples, where the flow was always downhill.

To understand this better,  let us explicitly see how this downhill flow is achieved in the Freund-Rubin case. The relevant equation is
\begin{aleq}
    \phi'' = \partial_\phi V -(d-1) \phi' A',
\end{aleq}
where the positive sign on the potential gradient reflects that the flow coordinate is spatial (unlike in cosmological evolution). The residual potential is negative and decays exponentially with the volume modulus,
\begin{aleq}
    V = -c^2 e^{-2\lambda \phi}.
\end{aleq}
This would lead to a positive (uphill) acceleration (\( \partial_\phi V > 0 \)) for the volume modulus. However, the friction term \( -(d-1) \phi' A' \) is negative and dominates, ensuring that \( \phi'' < 0 \). As a result, the flow remains downhill towards smaller volumes and vanishing metric factor \( e^A \to 0 \). A similar downhill behavior is found in all other pure flux vacua considered previously.

In contrast, for the KKLT scenario we find that \( \partial_\phi V < 0 \) for all moduli, and from the flow equations,
\begin{aleq}
-(d-1)\phi'A' < 0,
\end{aleq}
so that
\begin{aleq}
\phi'' < 0,
\end{aleq}
confirming that the flow is uphill towards smaller volumes and vanishing metric factor.

To understand better the flow near the singularity (and find the \emph{local} dynamical cobordism), we approximate
\begin{aleq}
    t(A) = c_1 e^{2A}, \quad s(A) = c_2 e^{2A}, \quad z(A) =c_3 e^{2A},
\end{aleq}
in which case the ten-dimensional metric is given by
\begin{aleq}
    ds_{10}^2 &= e^{2\phi}\mathcal{V}^{-1} \left( dr^2 + e^{2A(r)}ds_3^2\right) + \mathcal{V}^{1/3} ds_{CY}^2\\
    &=s^{-2}t^{-3/2}\left( dr^2 + e^{2A(r)}ds_3^2\right) +t^{1/2} ds_{CY}^2\\
    &= e^{-7A} dr^2 + e^{-5A} ds_3^2 + e^A ds_{CY}^2\\
    &= e^{(1+2p)A} dA^2 + e^{-5A} ds_3^2 + e^A ds_{CY}^2\\
    &= dy^2 + y^{-\frac{10}{1+2p}} ds_3^2 + y^{\frac{2}{1+2p}} ds_{CY}^2,
\end{aleq}
where we used that $dr \approx e^{(4+p)A}dA$ in this limit and defined $y\sim e^{(1+2p)A/2}$. We however remark that near the singularity, there might be significant corrections to the K\"ahler potential that could change the flow solution.
\subsection{Removing the \texorpdfstring{$F_3$}{F3}-flux}

As a toy model, the superpotential consists of
\begin{aleq}
W = (f_0+h_0s) + A_0 e^{-at}.
\end{aleq}

We remove the flux $f_0$ contribution to the potential
\begin{aleq}
W= h_0s+ A_0 e^{-at}, 
\end{aleq}
The flow equations are given by
\begin{aleq}
& t' =\frac{2 e^{-a t } z ^{-p/2} \left(A_0 (2 a t +3)+3 B s  e^{a t }\right)}{3 \sqrt{t  s }}, \quad
& s' = \frac{2 \sqrt{s } e^{-a t } z ^{-p/2} \left(A_0-B s  e^{a t }\right)}{t ^{3/2}},\\
& z' = \frac{2 e^{-a t } z ^{1-\frac{p}{2}} \left(B s  e^{a t }+A_0\right)}{\sqrt{t ^3 s }}, \quad
    & A' =\frac{z ^{-p/2} \left(A_0 e^{-a t }+B s \right)}{\sqrt{t ^3 s }}.
\end{aleq}
Now the evolution of the volume modulus and the dilaton as $A \rightarrow -\infty$ depends on the signs and magnitude of $A_0$ and $h_0$. When $A_0$ and $h_0$ have the same sign, the volume will be shrinking at the singularity (with small complex structure moduli as well) analogously to the case in subsection \ref{app:KKLT1}. The evolution of the dilaton will depend on the magnitudes of $A_0$ and $h_0$. If they have different signs, we could not find an obvious running solution in which the volume shrinks.

\bibliographystyle{utphys}  
\bibliography{references}
\end{document}